\documentclass[12pt]{article}

\def\bSig\mathbf{\Sigma}

\usepackage{graphicx}

 \usepackage{amssymb}
 \usepackage{amsmath}
 \usepackage{graphicx,caption}
 \usepackage{amsfonts}
 \usepackage{mathrsfs}
 \usepackage{verbatim}
 \usepackage[usenames, dvipsnames]{color}
 \usepackage[capposition=top]{floatrow}
 \usepackage[section]{placeins}
 \usepackage{rotating}
 \usepackage{multirow}
 \usepackage[colorlinks=true,linkcolor=blue,citecolor=blue]{hyperref}%
 \usepackage{lscape}
\usepackage{ulem}
\usepackage{xcolor}
 \usepackage{mathtools}
  \usepackage{lineno}
  \usepackage{bm}
  \usepackage{graphicx}
  \usepackage{graphicx,psfrag,epsf}
\usepackage{enumerate}
  \usepackage{natbib}
\graphicspath{ {./images/} }
  \newcommand{\Lower}[1]{\smash{\lower 1.5ex \hbox{#1}}}

  \newcommand{\blind}{0}

\def\bfb{{\ensuremath{\bf b}}}

\def\bfg{{\ensuremath{\bf g}}}
\def\bfh{{\ensuremath{\bf h}}}

\def\bfx{{\ensuremath{\bf x}}}

\def\bfA{{\ensuremath{\bf A}}}

\def\bfV{{\ensuremath{\bf V}}}

\def\bfX{{\ensuremath{\bf X}}}

\def\bfZ{{\ensuremath{\bf Z}}}
\def\bfzero{{\ensuremath{\bf 0}}}

\def\bftheta{{\ensuremath\boldsymbol{\theta}}}

\def\bfeta{{\ensuremath\boldsymbol{\eta}}}
\def\bfalpha{{\ensuremath\boldsymbol{\alpha}}}
\def\bfbeta{{\ensuremath\boldsymbol{\beta}}}
\def\bfgamma{{\ensuremath\boldsymbol{\gamma}}}

\def\bfrho{{\ensuremath{{\boldsymbol{\rho}}}}}

\def\bfPsi{{\ensuremath\boldsymbol{\Psi}}}
\def\bfPhi{{\ensuremath\boldsymbol{\Phi}}}

\def\bfmu{{\ensuremath{{\boldsymbol{\mu}}}}}

\def\bflambda{{\ensuremath\boldsymbol{\lambda}}}

  \newenvironment{eqarray*}{\arraycolsep 0.14em\begin{eqnarray*}}{\end{eqnarray*}}

  \numberwithin{equation}{section}
\usepackage{multirow}
\usepackage{graphicx}
\usepackage{booktabs}
\usepackage{xcolor}

\usepackage[utf8]{inputenc}
\usepackage{graphicx}
\usepackage{float}

\begin{document}

\def\spacingset#1{\renewcommand{\baselinestretch}%
{#1}\small\normalsize} \spacingset{1}

%%%%%%%%%%%%%%%%%%%%%%%%%%%%%%%%%%%%%%%%%%%%%%%%%%%%%%%%%%%%%%%%%%%%%%%%%%%%%%

\if0\blind
{
  \title{\bf {Advancing Information Integration through Empirical Likelihood: Selective Reviews and a New Idea }
   \author{Chixiang Chen$^{*}$\\
    Division of Biostatistics and Bioinformatics, University of Maryland\\
    School of Medicine\\
    Department of Neurosurgery, University of Maryland\\
    School of Medicine\\
    University of Maryland Institute for Health Computing, Bethesda\\
    and\\
    Jia Liang\\
    Department of Biostatistics, St. Jude Children's Research Hospital,\\ 
    Memphis, Tennessee, 38105.\\
    and\\
    Elynn Chen\\
    Department of Technology, Operations, and Statistics, \\
    New York University, New York, 10012. \\
    Ming Wang\\
    Department of Population and Quantitative Health Sciences, School of Medicine, \\
    Case Western Reserve University, Cleveland, Ohio 44106. \\
% E-mail address for correspondence
Contact Email $^{*}$: chixiang.chen@som.umaryland.edu}
}
  \maketitle
} \fi

\if1\blind
{
  \bigskip
  \bigskip
  \bigskip
  \begin{center}
    {\LARGE\bf Title}
\end{center}
  \medskip
} \fi

\bigskip
\begin{abstract}
Information integration plays a pivotal role in biomedical studies by facilitating the combination and analysis of independent datasets from multiple studies, thereby uncovering valuable insights that might otherwise remain obscured due to the limited sample size in individual studies. However, sharing raw data from independent studies presents significant challenges, primarily due to the need to safeguard sensitive participant information and the cumbersome paperwork involved in data sharing. In this article, we first provide a selective review of recent methodological developments in information integration via empirical likelihood, wherein only summary information is required, rather than the raw data. Following this, we introduce a new insight and a potentially promising framework that could broaden the application of information integration across a wider spectrum. Furthermore, this new framework offers computational convenience compared to classic empirical likelihood-based methods. We provide numerical evaluations to assess its performance and discuss various extensions in the end.
\end{abstract}

\noindent%
\vfill

\spacingset{1.45} % DON'T change the spacing!

\section{Introduction}
\label{sec:1}
% \red{MW: citation index needs to be updated}
Data integration plays a pivotal role in biomedical studies by allowing independent datasets from multiple studies to be combined and analyzed together, thereby unlocking a wealth of insights that would otherwise remain hidden due to the small sample size in an individual study \citep{Haidich2010Meta, Lapatas2015Data}.  By synthesizing information, we can improve estimation efficiency, enhance statistical significance, boost prediction accuracy, and detect small signals that may not be apparent when analyzing individual datasets \citep{qin2022selective}. The improved analysis will inform better decision-making in biomedical studies and promote personalized and precision medicine. 

However, sharing raw data from independent studies imposes substantial challenges, primarily due to the need to safeguard sensitive participant information and the cumbersome paperwork involved in data sharing \citep{Alfonso2017Data,Vepakomma2018Split}. In biomedical research, raw data frequently encompass detailed medical histories, genetic details, and other personally identifiable information, all subject to rigorous privacy regulations and ethical guidelines. Preserving privacy is fundamental to upholding trust and ethical standards within the research community \citep{Rothstein2010deidentification,Kisselburgh2022ethics}. Nevertheless, accomplishing this while simultaneously facilitating research utilizing big data across multiple studies poses a complex undertaking.

One of the most popular methods designed to integrate information from different studies without sharing the raw data is meta-analysis \citep{Haidich2010Meta}. These methods pool the published results of multiple similar scientific studies to produce an enhanced estimate without utilizing the raw individual data from each study \citep{Borenstein2021Introduction}. In recent years, many new methods have been developed to integrate summary information under the setting of more complex data structures and modeling strategies: some are based on frequentist inference, such as empirical likelihood-based estimators \citep{qin1994empirical,chatterjee2016constrained,han2019empirical,zhang2020generalized,sheng2022synthesizing,zhai2022data,chen2023efficient,liang2024integrative}, generalized-meta estimators \citep{Kundu2019Generalized}, communication-efficient distributed estimation \citep{Jordan2018Communication,Duan2022Heterogeneity,han2024privacy}, renewal estimation \citep{Luo2020Renewable,Luo2023Statistical} ; while others are based on Bayesian inference incorporating external information into informative priors \citep{Ibrahim2015power,Cheng2019Informing,Jiang2023Elastic}. The applications of information integration span over the generalized linear models, survival models, partial linear models, etc, and some methods have been applied to advance medical research, with findings published in esteemed journals, including Nature Medicine \citep{Jin2021Individual}. Notably, Qin and his colleagues have reviewed some of these works and summarized these into a unified framework of calibration \citep{qin2022selective}.

Given the pivotal role of calibration techniques in unifying numerous existing methods, this article focuses on reviewing empirical likelihood-based methods for information integration. These methods form a crucial foundation for many calibration approaches, and we provide selective reviews highlighting recent updates in this area. Additionally, we introduce a novel insight and a potentially promising framework that could expand the application of information integration across a broader spectrum, such as classic generalized linear model and semi-parametric causal inference. It is noteworthy that this new framework offers computational convenience and stability compared to classic empirical likelihood-based methods. Furthermore, it holds the potential to flexibly integrate existing techniques, such as density ratio models \citep{sheng2022synthesizing, cheng2023semiparametric, huang2023simultaneous} and penalized regressions \citep{zhai2022data,huang2023simultaneous}, and  address complex scenarios, such as heterogeneity in covariates/conditional outcome distributions between studies and the incorporation of information from multiple external sources.

The remainder of this article is organized as follows. Section \ref{sec:2} presents selective reviews of information integration methods utilizing empirical likelihood, providing detailed model specifications, algorithms, and recent updates to address data heterogeneity. Section \ref{sec:3} elaborates on the new idea and discusses its advantages in detail. Section \ref{sec:4} offers numerical evidence to illustrate and assess the new concept. Finally, Section \ref{sec:5} explores potential extensions and concludes the article.

\section{Information Integration via Empirical Likelihood}
\label{sec:2}
\subsection{Basic set-up}\label{Notation}
We first introduce the basic set-up for the internal and external studies that are widely adopted in the literature.  Let $n_1$ be the number of independent and identically distributed (i.i.d) subjects in the internal study, regarded as our main studied cohort. For each subject $i$ in the internal study, we have individual-level data denoted by ${Y_i,{\bfX}_i,\bfZ_i}$, where ${Y_i}$ is the outcome, ${\bfX}_i$ is the vector containing well-recognized covariates in the literature, and ${\bfZ}_i$ is the vector of extra covariates that are not available in the external study. The conditional density function of the outcome is denoted by $f_1(Y|{\bfX},\bfZ; \bfbeta_0)$, where $\bfbeta_0$ is the $p$-dimensional true parameter vector. In literature, one focus of the internal study is to model the conditional mean of the outcome ${Y}_i$, i.e., $E_1(Y|{\bfX},\bfZ; \bfbeta_0)$, with the conditional expectation taken with respect to the internal data.

\begin{figure}
    \centering
    \includegraphics[width=0.8\textwidth]{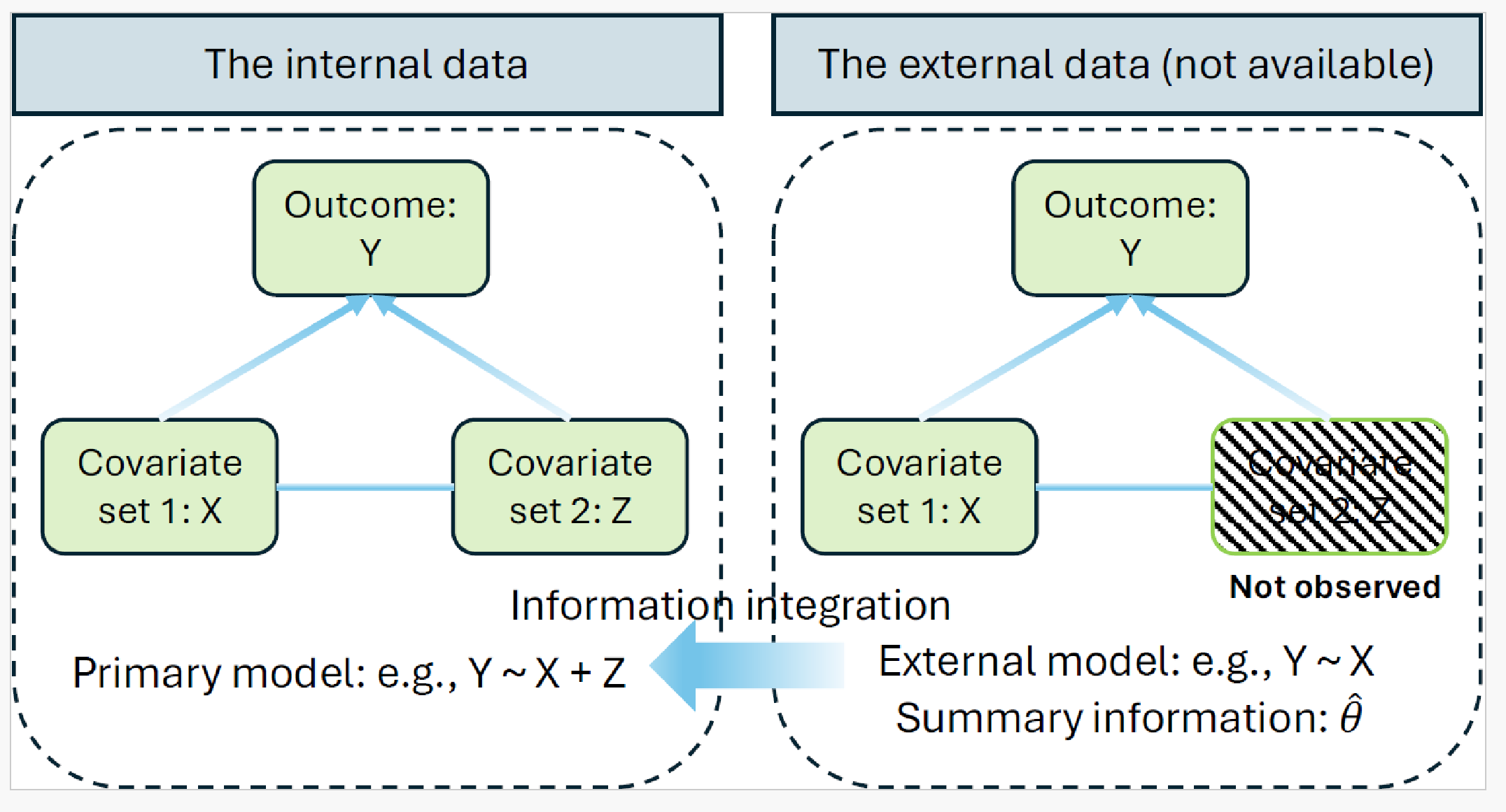}
    \caption{An illustrative example for data structure and method workflow in existing works.}
    \label{fig:1}
\end{figure}

On the other hand, let $n_2$ be the number of i.i.d subjects in the external study with the true conditional density function of the outcome denoted by $f_2(Y|\bfX,\bfZ;\bfbeta_0)$, which is often assumed to be the same to that from the internal study. For each subject $j$ from $1$ to $n_2$, only the outcome $Y_j$ and covariates $\bfX_j$ are assumed to be observed (Figure \ref{fig:1}). This is a reasonable and common setting in research where the external study has a large sample size but does not measure variables in $\bfZ_j$ from all participants, such as blood biomarkers, metabolic measures, or imaging metrics. These variables are often expensive to measure or are not considered in previous studies, but could be available in the internal study with a smaller sample size \citep{yang2019combining}. Additionally, suppose the raw data of the external study cannot be easily shared, while the summary information $\hat{\bftheta}$ could be available, which solves an estimating equation $\sum_{j=1}^{n_2}\bfPsi(Y_j,\bfX_j;\bftheta)=\bfzero$ based on the external data. Here, $\bfPsi(\cdot)$ can be any regular estimating function, for example, the score function based on a reduced and possibly mis-specified model $f_2(Y|{\bfX}; \bftheta_0)$ for the external data, with $\bftheta_0$ being the limiting values of $\bftheta$. The goal of information integration, therefore, is to utilize summary information $\hat{\bftheta}$ to enhance the precision of estimating $\bfbeta_0$ based on the internally studied data (Figure \ref{fig:1}), when the conditional density $f_1(Y|{\bfX},\bfZ; \bfbeta_0)$ is of the primary interest. A more general framework will be discussed in Section \ref{sec:3}. 
% \red{Tuoaa, do you think we can add a case where the estimating equations don't need to be the same internally and externally. This is like our causal scenario.}

\subsection{Constrained maximum likelihood}\label{Constrained Maximum Likelihood}

We start with describing the state-of-art method, named constrained maximum likelihood (CML) \citep{chatterjee2016constrained}. When the estimation variability of the summary information $\hat{\bftheta}$ can be ignored, the CML method provides a natural approach to leveraging this summary information to improve the inference of $\bfbeta$ \citep{chatterjee2016constrained, han2019empirical}. Specifically, CML is based on a semi-parametric likelihood where the density of the outcome given covariates is modeled by $f_1(Y|\bfX,\bfZ;\bfbeta)$, whereas the marginal density of covariates $f_1(\bfX,\bfZ)$ is modeled by an empirical distribution $p_i$ defined by the internal data. This distribution could be $1/n_1$, serving as a non-parametric estimate without incorporating any additional information. To integrate summary information $\hat\bftheta$ from the external study, the CML estimator $\hat\beta_{cml}$ is designed to maximize the following constrained optimization problem:
\begin{equation}\label{cml estimator}
    \sum_{i=1}^{n_1}\log f_1(Y_i|\bfX_i,\bfZ_i;\bfbeta)+\sum_{i=1}^{n_1}\log{p_i},
\end{equation}
 with respect to $p_i$ and is subject to three constraints
\begin{equation}\label{cml constraints}
p_i>0~, \forall i, ~\sum_{i=1}^{n_1}p_i=1, ~\sum_{i=1}^{n_1}p_i\bfPhi_1(\bfX_i,\bfZ_i;\bfbeta, \hat{\bftheta})=\bfzero,
\end{equation}
where $\bfPhi_1(\bfX,\bfZ;\bfbeta,\bftheta)=E_1\{\bfPsi(Y,\bfX;\bftheta)|\bfX,\bfZ;\bfbeta\}=\int\bfPsi(y,\bfX;\bftheta)f_1(y|\bfX,\bfZ;\bfbeta)dy$. 
% \red{Just a usual comment following lao Song's style, do you think we can use $dy$ rather than capitalized $Y$? (kind of like a random variable vs a fixed value in integration, may mean differently) And I guess the $Y$ under the integration sign indicates the domain of $Y$, right?} 

We notice here that the construction of $\bfPhi_1(\bfX,\bfZ;\bfbeta,\bftheta)$ links the internal main model and external reduced model by using the observed likelihood $f_1(Y|\bfX,\bfZ;\bfbeta)$, which makes the information integration feasible. Moreover, the expression in (\ref{cml estimator}) involves the logarithm of the joint likelihood of the data $\{Y_i,\bfX_i, \bfZ_i \}$, where the marginal density of covariates $\{\bfX_i, \bfZ_i\}$ remains unspecified {for (2)}. The following three constraints are then employed to identify the values of $p_i'$s, following the empirical likelihood philosophy \citep{qin1994empirical}. 

If we exclusively rely on the internal data to estimate $\bftheta$ and $\bfbeta$, we anticipate little efficiency gains compared to the well-established maximum likelihood (ML) estimator. This is understandable since the ML estimator optimally exploits the internal data, given regularity conditions \citep{daniels1961asymptotic}. However, the CML estimator is shown to be more efficient than the ML estimator by incorporating summary information $\hat\bftheta$ estimated from the external study, without estimating $\bftheta$ based on the internal data \citep{chatterjee2016constrained}. Intuitively, this additional information capitalizes on the extra degree of freedom from the moment condition $E_1\big\{\bfPsi(Y, \bfX;{\bftheta_0})\big\}=\bfzero$ 
% \red{This notation confuses me a bit by reading through. Since we previously mentioned $\hat\bftheta$, now it directly jumps to $\bftheta_0$. Maybe we can add a sentence in between?}
, leading to a more efficient estimate of $\bfbeta$ compared to the ML estimator.

We also remark here that the CML method considers the setup where the variability of $\hat\bftheta$ can be ignored. This is the case when the sample size $n_2$ of the external study is much larger than the sample size $n_1$ of the internal study. However, when $n_2$ is comparable or even smaller than $n_1$, the CML estimator will underestimate the variance \citep{han2019empirical,zhang2020generalized}. Moreover, the construction of $\bfPhi_1(\bfX,\bfZ;\bfbeta,\bftheta)=E_1\{\bfPsi(Y,\bfX;\bftheta)|\bfX,\bfZ;\bfbeta\}$ and the use of covariate probability mass $\hat p_i$ imply that the marginal distributions of covariates $\{\bfX,\bfZ\}$ between internal and external studies should be the same in general to ensure the consistent estimate of $\bfbeta_0$. 

\subsection{Generalized integration model}\label{Generalized integration model}
When the estimation variability of $\hat\bftheta$ cannot be ignored, a more general and possibly more practical scenario 
% \red{scenario or model? If "scenario", should we say "a more general and possibly more practical scenario in application demands the GIM (cite Dr. Qin's paper)"} 
in real applications, the generalized integration model (GIM) should be adopted \cite{zhang2020generalized}. Specifically, GIM extended the CML estimation by adding an extra penalty term to the constrained optimization, accounting for the estimation variability of $\hat{\bftheta}$. 
% This leads to a new constrained optimization problem that can be maximized to obtain the updated estimator $\hat\bfbeta_{gim}$: 
{This leads to a new constrained optimization problem, which maximizes the following likelihood to obtain the estimator $\hat\bfbeta_{gim}$: }
\begin{equation}\label{gim estimation}
    \sum_{i=1}^{n_1}\log f_1(Y_i|\bfX_i,\bfZ_i;\bfbeta)+\sum_{i=1}^{n_1}\log{p_i}-n_2(\hat{\bftheta}-\bftheta)^T\hat\bfV^{-1}(\hat{\bftheta}-\bftheta)/2,
\end{equation}
 with respect to $p_i$ and $\bftheta$ and is subject to three constraints:
\begin{equation}\label{gim constraints}
p_i>0,~~\forall i,~~\sum_{i=1}^{n_1}p_i=1,~~\sum_{i=1}^{n_1}p_i\bfPhi_1(\bfX_i,\bfZ_i;\bfbeta,{\bftheta})=\bfzero,
\end{equation}
where $\hat\bfV$ is a consistent estimate of the variance-covariance matrix of $n_2^{0.5}(\hat{\bftheta}-{\bftheta}_0)$. 
% \red{what about just $\hat\bfV$ and $\bfV_0$, $\theta$ and $\beta$ follows those scheme.}
 Intuitively, the expression in (\ref{gim estimation}) can be regarded as the logarithm of the joint distribution of the data $\{Y_i,\bfX_i,\bfZ_i,\hat\bftheta\}$, for $i=1,\ldots,n_1$. Therefore, the uncertainty of $\hat\bftheta$ is naturally incorporated into the information integration by assuming that $n_2^{0.5}(\hat{\bftheta}-{\bftheta}_0)$ follows a multivariate normal distribution. This estimator is more efficient than both the ML estimator and the CML estimator, and it reaches the semi-parametric efficiency bound \citep{zhang2020generalized}. The developed framework also allows the estimation of nuisance parameters, such as the over-dispersion parameter in generalized linear models \citep{zhang2020generalized}. 

Despite advancements, GIM still relies on the assumption that the marginal distributions of covariates are consistent across the two studies. 
%\blue{GIM is not about internal and multiple externals, right? Or they have discussed it as well?} \red{they provided extension to multiple ones, but not evaluated}

\subsection{The numerical procedure}\label{Algorithms}
To successfully deliver information, the parameters including $p_i$, $\bfbeta$, and $\bftheta$ (only in GIM) should be estimated simultaneously in a constrained optimization procedure. The typical numerical procedure involves the method of Lagrange multipliers to profile out the empirical distribution parameters \cite{qin1994empirical,chatterjee2016constrained, zhang2020generalized}. Let us take the estimation of GIM for illustration. The Lagrange function will be defined as 
\begin{equation}\label{gim algorithm}
\begin{split}
        L_{n_1}(p_1,\ldots,p_{n_1},\bfbeta,\bftheta,\bflambda, t)=&\sum_{i=1}^{n_1}\log p_i+\sum_{i=1}^{n_1}\log f_1(Y_i|\bfX_i,\bfZ_i;\bfbeta) \\
        &-n_2(\hat\bftheta-\bftheta)^T\hat\bfV^{-1}(\hat\bftheta-\bftheta)/2
        -n_1t\bigg(\sum_{i=1}^{n_1}p_i-1\bigg)\\
        &-n_1\sum_{i=1}^{n_1}p_i\bflambda^T\bfPhi_1(\bfX_i,\bfZ_i;\bfbeta,\bftheta),
\end{split}
\end{equation}
with $(\bflambda^T,t)^T$ being the Lagrange multipliers. Solving the above function using the constraints $\sum_{i=1}^{n_1}p_i=1$ and $\sum_{i=1}^{n_1}p_i\bflambda^T\bfPhi_1(\bfX_i,\bfZ_i;\bfbeta,\bftheta)=\bfzero$, we have $t=1$ and $p_i=n_1^{-1}\big\{1+\bflambda^T\bfPhi_1(\bfX_i,\bfZ_i;\bfbeta,\bftheta)\big\}^{-1}$. Thus, the function in (\ref{gim algorithm}) will be reduced to 
\begin{equation}\label{gim algorithm reduce}
\begin{split}
        l_{n_1}(\bfbeta,\bftheta,\bflambda)=&-\sum_{i=1}^{n_1}\log \big\{1+\bflambda^T\bfPhi_1(\bfX_i,\bfZ_i;\bfbeta,\bftheta)\big\}+\sum_{i=1}^{n_1}\log f_1(Y_i|\bfX_i,\bfZ_i;\bfbeta) \\
        &-n_2(\hat\bftheta-\bftheta)^T\hat\bfV^{-1}(\hat\bftheta-\bftheta)/2.
\end{split}
\end{equation}
As a result, solving (\ref{gim algorithm reduce}) is translated into solving the unconstrained optimization: $\max_{\bfbeta,\bftheta,\bflambda}l_{n_1}(\bfbeta,\bftheta,\bflambda)$.

To obtain the estimates of $\bfbeta,\bftheta,\bflambda$, we need to solve the following estimating equations by an iterative manner:
\begin{equation*}
\begin{split}
        \frac{\partial l_{n_1}(\bfmu,\bflambda)}{\partial\bfmu}=\bfzero,~\frac{\partial l_{n_1}(\bfmu,\bflambda)}{\partial\bflambda}=\bfzero,
\end{split}
\end{equation*}
with $\bfmu=(\bfbeta^T,\bftheta^T)^T$. Note that given the values of $\bfmu$, the second estimating function is convex,  which can be efficiently solved by greedy algorithms \citep{chen2008adjusted,han2014further,han2019empirical}. Given the values of $\bflambda$, the first equation can be solved by Newton-Raphson method \citep{han2019empirical,zhang2020generalized}. 
% Always give a unique label
% and use \ref{<label>} for cross-references
% and \cite{<label>} for bibliographic references
% use \sectionmark{}
% to alter or adjust the section heading in the running head

\subsection{Various extensions}\label{Various extensions}
The CML estimator and the GIM estimator establish the theoretical foundation that permits various methodological extensions. We provide a brief overview of several variants designed to tackle the following three challenges: heterogeneous covariate distributions between two studied cohorts, heterogeneous conditional outcome distributions between two studied cohorts, and incorporating summary information from multiple external studies.

\textbf{Heterogeneous covariate distributions}. The CML and GIM procedures assume homogeneity in the covariate distribution. Given variations in inclusion and exclusion criteria across studies, this assumption may not hold, potentially resulting in biased estimates. To address heterogeneous covariate distributions, one may consider adopting a semiparametric density ratio model \citep{sheng2022synthesizing,cheng2023semiparametric,huang2023simultaneous}, i.e., assume two density functions satisfy the following relationship
\begin{equation}\label{density ratio model}
f_2(\tilde\bfX)=\exp(\alpha_0+\tilde\bfX^T\bfalpha)f_1(\tilde\bfX),
\end{equation}
where $f_1(\tilde\bfX)$ and $f_2(\tilde\bfX)$ denote the density functions of $\tilde\bfX$ in the internal and external studies, respectively, with $\tilde\bfX$ being a subset vector of $\bfX$ and unknown parameters in $\bfalpha$. The scalar $\alpha_0$ normalizes the function such that $\int f_2(\tilde\bfx)d\tilde\bfx=1$.
% \red{(Tuoaa, do you think we should emphasize the distribution of $\tilde\bfX$ inside the exponential term as well? Because when I do not know the semiparametric density ratio model, I would be guessing if this $\tilde\bfX$ on the RHS comes from external, but not sure.)}
% \blue{not quite understand what you mean} 
Based on the above density ratio model, the CML and GIM can be easily modified by adding an additional constraint, i.e., $\sum_{i=1}^{n_2}p_i\big\{\exp(\tilde\bfX^T\bfalpha)-1\big\}=0$, into (\ref{cml constraints}) and (\ref{gim constraints}). The rationale of adding this constraint is based upon the fact that $E_1\big\{\exp(\tilde\bfX^T\bfalpha)\bfPhi_1(\bfX_i,\bfZ_i;\bfbeta,{\bftheta})\big\}=\bfzero$. When $\bfalpha=\bfzero$, the method will be reduced CML/GIM estimation. In general case, the parameter vector $\bfalpha$ is unknown and needs to be estimated.
% \red{(this one? $\int_{I_{(\tilde\bfX)}}f_{2}(t)dt=1$)}. 
By introducing and jointly estimating extra parameter vector $\bfalpha$, the estimators derived from the modified constrained optimization is able to calibrate covariate distribution difference and thus unbiased if the density ratio model is correctly specified. This technique has found application in various contexts, including survival data analysis \citep{cheng2023semiparametric}. Two important notes are highlighted: due to the limitedly available external data in summary forms, \textbf{(1)} assessing whether the specification of density ratio model in (\ref{density ratio model}) is correct or not is challenging or even impossible; \textbf{(2)} Even in the case the density ratio model holds, the identification of the subvector $\tilde\bfX$ is also challenging \citep{huang2023simultaneous}. In practice, researchers may consider some important covariates, such as race/ethnicity, that are believed to be different between datasets \citep{sheng2022synthesizing}. 

\textbf{Heterogeneous conditional outcome distributions}. In addition to covariate distributions, heterogeneous conditional outcome distributions, $f_1(Y|\bfX,\bfZ)\neq f_2(Y|\bfX,\bfZ)$, will also lead to biased estimators and may impose more challenge to calibrate. One promising solution is to introduce a bias term $\bfb$ into the constraint \citep{zhai2022data,huang2023simultaneous}, i.e., 
\begin{equation}\label{bias term}
    \sum_{i=1}^{n_1}p_i\bfPhi_1(\bfX_i,\bfZ_i;\bfbeta,{\bftheta})-\bfb=\bfzero.
\end{equation}
Intuitively, the vector $\bfb$ models the values 
% \red{(I guess it's the tuples in the vector that deviate from 0?)} 
of the underlying moment $E_1\big\{\bfPsi(Y, \bfX_i;{\bftheta_0})\big\}$, where some elements could deviate from zero when conditional outcome distributions are different.
% , where $\bftheta_0$ represents the limiting value of $\hat\bftheta$ from the external study, ensuring unbiased estimation of $\bfbeta$. 
We notice here that the values in $\bfb$ are unknown and need to be estimated using the internal data. Without imposing an extra constraint on estimating the bias term $\bfb$, there would be no efficiency gain for estimating $\bfbeta$. To facilitate information integration, an $l_1$-based penalty is suggested to be incorporated into the constrained optimization process, effectively estimating $\bfb$ and shrinking the values in $\hat\bfb$ that are close to zero. The idea using penalty is to shrink the estimated elements of $E_1\big\{\bfPsi(Y, \bfX_i;{\bftheta_0})\big\}$ that are truly zero and leave the other elements unshrinked. Consequently, the information delivery can be still achievable via the elements $\hat{\bfb}_\ast=\bfzero$ with $\hat{\bfb}_\ast\subset \hat\bfb$. For a more detailed description of the estimation procedure, we direct readers to the relevant literature \citep{zhai2022data,huang2023simultaneous}.

\textbf{Multiple external studies}. Integrating summary information from multiple external studies can be instrumental in further boosting estimation efficiency for $\bfbeta$. Under the assumption of homogeneous populations, the GIM estimator is able to seamlessly incorporates multiple external estimates $\hat\bftheta_k$ for $k=2,\ldots,K$ and $K\geq 2$. This modification is achieved by adjusting the last term in the expression in (\ref{gim estimation}) to $\sum_{k=2}^Kn_k(\hat{\bftheta}_k-\bftheta)^T\hat\bfV_k^{-1}(\hat{\bftheta}_k-\bftheta)/2$, where $\hat\bfV_k$ is a consistent estimate of the covariance matrix of $n_k^{0.5}(\hat\bftheta_k-\bftheta_0)$, with the sample size of $n_k$ in the k-th study. 

In addition, in the complex situation where multiple external datasets are believed to exhibit different covariate distributions and conditional outcome distributions, one can adapt techniques from the density ratio model and bias penalty, as described earlier, to facilitate the integration of information from multiple external studies \citep{huang2023simultaneous}.

\section{A New Idea}
\label{sec:3}
\subsection{Method Framework}\label{Method Framework}
Despite theoretical advancements, the methods described above involve the conditional distribution $f_1(Y|\bfX,\bfZ;
\bfbeta)$ to link models from internal and external studies, which may not always be applicable to general semi-parametric estimation. Moreover, these methods often necessitate complex computational strategies to jointly estimate all parameters, as described in Section \ref{Algorithms}. In this section, we introduce a new perspective and a general framework for information integration that has a significantly lighter computational load and encompasses a possibly broader range of application contexts.

Before describing the proposed method, let us refine the previous notations and consider a broader context. Suppose the quantity of interest $\bfbeta$ can be identified by a generic estimating equation $E_1\big\{\bfg(Y,\bfX,\bfZ;\bfbeta_0,\bfeta_0)\big\}=\bfzero$, with a vector $\bfeta_0$ consisting of the true values of potential nuisance parameters and an i.i.d estimation function $\bfg(Y_i,\bfX_i,\bfZ_i;\bfbeta_0,\bfeta_0)$, for $i=1,\ldots,n_1$. We remark here that the above setting does not require the full specification of the observed likelihood, and the interested parameters in $\bfbeta$ are not limited to the parameters in the conditional outcome distribution.
% , making it more general to encompass a broad range of statistical modeling strategies. 
Two examples, but not limited to two, are illustrated below:

\textbf{Example 1: Generalized linear model (GLM)}. In the GLM setting, $\bfg(Y,\bfX,\bfZ;\bfbeta,\bfeta)$ could be the score function with the parameter vector $\bfbeta$ indexed in the conditional mean structure $\mu(\bfX,\bfZ;\bfbeta)$, i.e., $\bfg(Y,\bfX,\bfZ;\bfbeta,\bfeta)=(1,\bfX^T,\bfZ^T)^T\big\{Y-\mu(\bfX,\bfZ;\bfbeta)\big\}$. Under this GLM framework, CML and GIM estimation procedures are still applicable to facilitate information integration (Figure \ref{fig:1}).  

\textbf{Example 2: Causal inference}. Unlike GLM, which allows us directly work on the observed likelihood, causal inference models often rely on the potential outcome framework with calibrated moment conditions \citep{rosenbaum1983central,austin2015moving,chen2024multiple}. For instance, consider our interest in the marginal and causal odds ratio between two groups (e.g., $A=0,1$, a scalar), and let the vector $\bfX$ defined before contain the the exposure variable $A$, i.e., $\bfX=(A,\bfX_\ast^T)^T$ with a sub-vector $\bfX_\ast$ of $\bfX$ excluding the variable $A$.
% \red{HERE, originally it was }. 
In this context, an unbiased estimator could be identified based on the moment condition $E_1\big\{\bfg(Y,\bfX_{{\ast}},\bfZ;\bfbeta_0,\bfeta_0)\big\}=\bfzero$ with:
\begin{equation*}
\bfg(Y,\bfX,\bfZ;\bfbeta,\bfeta)=
    \begin{pmatrix}
\frac{1}{\pi(\bfX_\ast,\bfZ;\bfeta)}(1,A)^T\big\{Y-\mu(\bfA;\bfbeta)\big\}\\
\bfh(\bfX,\bfZ;\bfeta)
\end{pmatrix}.
\end{equation*}
Here, $\mu(A;\bfbeta)=\big\{1+\exp(-\beta_0-\beta_1A)\big\}^{-1}$; the quantity $\pi(\bfX_{\ast},\bfZ;\bfeta)$, so-called propensity score (PS), equals the probability of $A$ given the covariates and serves as a calibration weight to balance the confounder distributions between two groups \citep{rosenbaum1983central,austin2015moving,chen2024multiple}; $\bfh(\bfX,\bfZ;\bfeta)$ is an estimating function solving the parameters in $\bfeta$, which could be the score function from logistic regression by treating $A$ as the outcome. Stacking two estimating functions together aim to create an i.i.d estimating function $\bfg(Y,\bfX,\bfZ;\bfbeta,\bfeta)$ to ensure efficiency gain in theory \citep{liang2024integrative}.  Thus, the parameter of interest in this case is the causal odds ratio $\beta_1$ based upon the marginal structural model (MSM) under the potential outcome framework \citep{robins2000marginal}. It is important to note that the CLM and GIM estimators may not be directly applicable in this case for integrating information, as the observed likelihood may no longer represent the distribution of pseudo outcomes in the causal context. Therefore, a new integration method is needed.

To integrate the information from the summary information $\hat\bftheta$ under the above setting, we propose the following weighted estimation procedure: the estimator of $\bfbeta$ and $\bfeta$ (if exist) could be jointly obtained by solving the weighted estimating equation:
\begin{equation}\label{our estimatator}
\sum_{i=1}^{n_1}\hat p_i\bfg(Y_i,\bfX_i,\bfZ_i;\bfbeta,\bfeta)=\bfzero,
\end{equation}
where the weight $\hat p_i$ is estimated by maximizing the {joint log-likelihood} $l$, 
\begin{equation}\label{Vtheta}
    l=\sum_{i=1}^{n_1} log(p_i) - n_2(\hat\bftheta-{\bftheta})^T({\hat\bfV})^{-1} (\hat\bftheta-{\bftheta})/2,
\end{equation}
 with respect to $p_i$ and $\bftheta$, and is subject to three constraints:
\begin{equation}\label{constrain with variance}
    p_i>0~, \forall i; ~~\sum_{i=1}^{n_1} p_i=1, ~~\sum_{i=1}^{n_1} p_i\bfPsi(Y_i, {\bfX}_{i};\bftheta)=\bfzero.
\end{equation}

Intuitively, the construction of the weight $\hat p_i$ relies on a semi-parametric joint {log-}likelihood of the internal data $\{Y,{\bfX}\}$ and the summary information {$\hat\bftheta$} from the external study, where the external information is delivered through the moment constraint in (\ref{constrain with variance}), and the estimation uncertainty is accounted by the quadratic term in (\ref{Vtheta}), analogous to the rationale in the GIM estimator. The external information is expected to be integrated by the constructed semi-parametric joint {log-}likelihood, where the weight $\hat p_i$ is a more efficient estimate of empirical distribution \citep{qin1994empirical}. Thus, the resulting $\hat p_i$ serves an informative weight, carrying additional information from the external data and integrating it into the internal estimating equation \citep{chen2022improving, chen2023efficient}.  

Distinct from the GIM procedure, however, the estimation of main parameters in $\bfbeta$ is not involved in the estimation of $p_i$ and $\bftheta$. This feature decouples the estimation of the main parameter vector $\bfbeta$ and extra parameters $p_i$ and $\bftheta$ for information integration, which reduces the computational load in comparison to the joint estimation algorithm described in Section \ref{Algorithms}. More importantly, the above procedure avoids the specification of observed likelihood to link model systems between two studies, i.e., $\sum_{i=1}^{n_1}\bfg(Y_i,\bfX_i,\bfZ_i;\bfbeta,\bfeta)=\bfzero$ and $\sum_{i=1}^{n_1}\bfPsi(Y_i, {\bfX}_{ i};\bftheta)=\bfzero$. 
% \red{I think to understand the first half of the sentence, one has to have already read the GIM paper or as you have already described it to me. What about "This avoids the knowledge of specific likelihood that satisfies both studies,  which is needed to borrow information. And, it allows moderate heterogeneous covariate distribution"} Instead, we only require a milder moment condition: $E_1\big\{\bfPsi(Y, \bfX_i;{\bftheta_0})\big\}=\bfzero$, with $\bftheta_0$ being the limiting value of $\hat\bftheta$ from the external study. This moment condition holds if conditional outcome distributions are the same, regardless of marginal covariate distributions. 
Therefore, the proposed integration framework has larger potential to accommodate broader modeling strategies, particularly in cases where deriving an analytical relationship between internal and external models may prove challenging.

We remark here that the proposed framework has recently been employed to integrate summary information, enhancing estimation efficiency in a partial linear model (PLM) \citep{liang2024integrative}. The theoretical framework has demonstrated that the resulting estimator is more efficient than the typical profile least square estimator \citep{fan2004new}. This article extends its application to a more generic setting, including GLM and causal estimation, and we anticipate the same asymptotic properties by following the proof strategy outlined in that literature \citep{liang2024integrative}, omitting the detailed proof here. In the subsequent sections of this article, we describe the computational advancements and provide numerical evidence through simulation studies to evaluate the validity of the proposed framework in terms of estimation bias and Monte Carlo standard deviation. Furthermore, we discuss potential extensions of this new framework to handle complex situations as described in Section \ref{Various extensions}, such as heterogeneous covariate distributions, heterogeneous conditional outcome distributions, and information integration from multiple external studies.

% Always give a unique label
% and use \ref{<label>} for cross-references
% and \cite{<label>} for bibliographic references
% use \sectionmark{}
% to alter or adjust the section heading in the running head

\subsection{Efficient Computation}\label{Efficient Computation}

 As described from the last section, the proposed estimation requires much less computational load, in comparison to CML and GIM-based estimators. The decoupled feature leads to more efficient computational strategy to estimate the main parameters in $\bfbeta$ by solving using weighting estimation equation. However, the estimation of weight is still entangled with $\bftheta$ estimation, which requires iterative updating algorithm from the empirical likelihood framework and could be time-consuming as well. To further release computational load, we advocate the following estimation procedure: 
 
 \textbf{Step 1}: Estimate $\bftheta$ via the meta analysis, i.e., by minimizing $\sum_{i\in \{0,1\}}(\bftheta-{\hat\bftheta_{(i)}})^T{\hat\bfV_{(i)}^{-1}}(\bftheta-{\hat\bftheta_{(i)}})$, where ${\hat\bftheta_{(1)}}={\hat\bftheta}$ and $
{\hat\bfV_{(1)}}=\hat \bfV/n_2$; ${\hat\bftheta_{(0)}}$ is the estimate solved by the estimating equation $\sum_{i=1}^{n_1}\bfPsi(Y_i, {\bfX}_i;\bftheta)=\bfzero$ based on the internal data, and ${\hat\bfV_{(0)}}$ is a {consistent estimate of the }variance-covariance matrix of ${\hat\bftheta_{(0)}}$. We denote the resulting estimate as $\hat\bftheta_{meta}$. 

 \textbf{Step 2}: Using the plug-in estimator from the meta-analysis $\hat\bftheta_{meta}$, the weight $\hat p_i=(1/n_1)\big\{1+\hat\bfrho^T\bfPsi(Y_i, {\bfX}_i;\hat\bftheta_{meta})\big\}^{-1}$ is readily calculated, where the estimated Lagrange multiplier $\hat\bfrho$ is obtained by solving $\partial\tilde l_{n_1}(\bfrho)/\partial\bfrho=\bfzero$, with
\begin{equation}\label{easier loss}
\begin{split}
        \tilde l_{n_1}(\bfrho)=&-\sum_{i=1}^{n_1}\log \big\{1+\bfrho^T\bfPsi(Y_i, {\bfX}_i;\hat\bftheta_{meta})\big\}.
\end{split}
\end{equation}
Since $\tilde l_{n_1}(\bfrho)$ is a convex function in terms of $\bfrho$, the estimation of $\bfrho$ is quick and stable \citep{chen2008adjusted,han2014further,han2019empirical}. Two steps above are performed only once, without the need for iterative updating. 

The above two-step estimation has been demonstrated to be asymptotically equivalent to the empirical likelihood-based estimator by solving the constrained optimization in (\ref{Vtheta}) and (\ref{constrain with variance}) under the PLM setting \citep{liang2024integrative}. We expect a similar property in the context discussed in this paper and defer the detailed proof to interested readers.

\section{Numerical Evidence}
\label{sec:4}

\subsection{Generalized linear model}\label{generalized linear model}
\textbf{Data generation}. 
We considered binary outcome for illustration. Both internal and external data were generated through the model $\xi\{Prob(Y_i=1|\bfX_i, Z_i)\}=\beta_0+\beta_1X_{1i}+\beta_2X_{2i}+\beta_3X_{3i}+\beta_4 Z_i$ with a logit link function $\xi(\cdot)$. To consider potential confounding between covariates, we generated the covariate $Z_i$ following a uniform distribution from $0$ to $1$, the covariate $X_{1i}$ following a normal distribution with mean equal to $Z_i$ and variance equal to 1, the covariate $X_{2i}$ following a Bernoulli distribution with the probability $exp(Z_i)/\{1+exp(Z_i)\}$, and the covariate $X_{3i}$ following a standard normal distribution. The parameter vector was set to be $\bfbeta=(\beta_0,\beta_1, \beta_2, \beta_3, \beta_4)^T=(1,1,1,1,1)^T$. We tested two internal sample sizes $n_1=200,600$. The external sample size was set to be proportional to the internal sample size, where $r=n_2/n_1$ equals $0.75$, $1.5$, and $5$.

Following the setting described in Section \ref{Notation}, we considered the scenario where the variable $Z_i$ was not observed in the external data. Consequently, we assumed that the external study adopted the following model (possibly mis-specified): $\xi\{Prob(Y_i=1|\bfX_i)\}=\theta_0+\theta_1X_{1i}+\theta_2X_{2i}+\theta_3X_{3i}$. The estimated vector of parameters $\hat\bftheta$ and its estimated variance-covariance matrix $\hat\bfV$ were assumed to be available, but not the raw external data. We evaluated the performance of the new method, denoted by IB\_New,  and compared it with GLM and GIM estimators in terms of bias, Monte Carlo Standard Deviation (MCSD), and Relative Efficiency (RE), which is the ratio of mean square errors between the GLM estimator and an estimator with data integration (named New\_RE for the proposed estimator and GIM\_RE for the GIM estimator). All evaluations were conducted in R software under version 4.3.2.

\begin{table}[]
\caption{Comparison of parameter estimation via the proposed new method, the GIM, and the GLM approach}\label{glm_gim_table}
\centering
\renewcommand{\arraystretch}{1.5} 
\begin{tabular}{p{1cm}p{1cm}p{1.2cm}p{1.2cm}p{1.2cm}p{0.9cm}p{1.2cm}p{0.9cm}p{0.9cm}p{0.9cm}p{0.9cm}p{0.9cm}}
\hline
                       & \multicolumn{2}{c}{Ratio of $n_2/n_1$} & \multicolumn{3}{c}{0.75}       & \multicolumn{3}{c}{1.5}        & \multicolumn{3}{c}{5}          \\ \hline
                       &                 &                &   $~~\beta_1$ & $~~\beta_2$ & $~~\beta_3$ & $~~\beta_1$ & $~~\beta_2$ & $~~\beta_3$ & $~~\beta_1$ & $~~\beta_2$ & $~~\beta_3$ \\ \hline
\multirow{8}{*}{$n_1$=200} & Bias          & IB\_New         & 0.013    & 0.031    & 0.015    & 0.017    & 0.008    & 0.015    & 0.018    & 0.016    & 0.021    \\ %\cline{2-12} 
                       & Bias            & GLM            & 0.078    & 0.082    & 0.081    & 0.056    & 0.062    & 0.063    & 0.067    & 0.057    & 0.053    \\ %\cline{2-12} 
                       & Bias         & IB\_GIM        & 0.024    & 0.040    & 0.026    & 0.023    & 0.016    & 0.022    & 0.016    & 0.015    & 0.016    \\ %\cline{2-12} 
                       & MCSD          & IB\_New         & 0.226    & 0.399    & 0.231    & 0.190    & 0.338    & 0.191    & 0.136    & 0.226    & 0.127    \\ %\cline{2-12} 
                       & MCSD             & GLM            & 0.316    & 0.546    & 0.327    & 0.306    & 0.519    & 0.311    & 0.315    & 0.543    & 0.308    \\ %\cline{2-12} 
                       & MCSD         & IB\_GIM        & 0.223    & 0.392    & 0.228    & 0.188    & 0.333    & 0.190    & 0.131    & 0.213    & 0.122    \\ %\cline{2-12} 
                       &                 & New\_RE         & 2.073    & 1.899    & 2.125    & 2.657    & 2.387    & 2.739    & 5.523    & 5.799    & 5.900    \\ %\cline{2-12} 
                       &                 & GIM\_RE        & 2.115    & 1.964    & 2.165    & 2.717    & 2.462    & 2.732    & 5.996    & 6.565    & 6.434    \\ \hline
\multicolumn{12}{l}{}                                                                                                                                        \\ \hline

\multirow{8}{*}{$n_1$=600} & Bias          & IB\_New           & -0.002   & -0.001   & 0.002    & -0.000   & 0.000    & 0.008    & 0.005    & 0.011    & 0.009    \\ %\cline{2-12} 
                       & Bias             & GLM           & 0.020    & 0.023    & 0.022    & 0.015    & 0.006    & 0.027    & 0.027    & 0.049    & 0.033    \\ %\cline{2-12} 
                       & Bias         & IB\_GIM           & 0.001    & 0.001    & 0.005    & 0.003    & 0.002    & 0.011    & 0.006    & 0.014    & 0.010    \\ %\cline{2-12} 
                       & MCSD          & IB\_New           & 0.116    & 0.213    & 0.123    & 0.103    & 0.188    & 0.108    & 0.073    & 0.130    & 0.073    \\ %\cline{2-12} 
                       & MCSD             & GLM           & 0.158    & 0.275    & 0.166    & 0.162    & 0.280    & 0.164    & 0.161    & 0.290    & 0.167    \\ %\cline{2-12} 
                       & MCSD         & IB\_GIM           & 0.116    & 0.213    & 0.121    & 0.101    & 0.186    & 0.106    & 0.071    & 0.123    & 0.070    \\ %\cline{2-12} 
                       &                 & New\_RE         & 1.885    & 1.679    & 1.836    & 2.499    & 2.212    & 2.365    & 4.943    & 5.063    & 5.394    \\ %\cline{2-12} 
                       &                 & GIM\_RE        & 1.905    & 1.687    & 1.890    & 2.574    & 2.280    & 2.423    & 5.321    & 5.619    & 5.843    \\ \hline

\end{tabular}
\end{table}

\textbf{Results}. The results based on $1000$ Monte Carlo runs are summarized in Table \ref{glm_gim_table}. Both IB\_New and IB\_GIM exhibited minimal (even smaller) estimation bias and demonstrated a significant reduction in estimation variability (MCSD) compared to the GLM estimator without information integration across all settings. It is also interesting to note that the IB\_New estimator performed very similarly to IB\_GIM in terms of RE. These findings suggest that both GIM and the proposed new method effectively integrate information to enhance the analysis of internal studies.

However, the computational difference between GIM and the proposed estimation is significant. The average running time for GIM on one Monte Carlo simulation with an internal sample size of $600$ was $27$ seconds, while the average time for our new method was only $1.8$ seconds, making it ten times more efficient than the GIM method. Moreover, the GIM method was observed to be more prone to algorithm convergence issues with smaller sample sizes. Therefore, the new method offers a significant computational advantage, which is highly valuable in practice.

% The results based on $1000$ Monte Carlo runs are summarized in Table \ref{glm_gim_table}. Both IB\_New and IB\_GIM led to little (even smaller) estimation bias and showed significant reduction of estimation variability (MCSD), compared to the GLM estimator without information integration in all settings. It is also interesting to observe that the IB\_New estimator had very similar performance compared to the IB\_GIM in terms of RE. These results imply that both GIM and the proposed new method are able to integrate information to enhance the internal study analysis. However, {the average running time used by GIM on one Monte Carlo simulation with internal sample size $600$ was 27 seconds, while the average time used by our new method was 1.8 seconds, which was ten times more efficient than the GIM method. Moreover, the GIM method was observed to be more likely to have algorithm convergence issue when sample size was smaller.} Thus, the new method holds computational advantage, highly valuable in practice.  

% then used in the empirical likelihood (\ref{our estimatator}), and (\ref{Vtheta}) to solve for the informative score $p_i$. See table \ref{glm_gim_table} for simulation results after $1000$ Monte Carlo runs. 

\subsection{Causal inference model}\label{Casual inference model}

\textbf{Data generation}. 
In this section, we generated the data under the context of causal inference with a binary outcome and a binary exposure. Specifically, for both internal and external data, the exposure was generated based on the Bernoulli distribution, where the probability of success was set to be $\xi\{Prob(A_i=1|\bfX_i, Z_i)\}=\gamma_0+\gamma_1Z_i+\gamma_2X_{1i}+\gamma_3X_{2i}+\gamma_4X_{3i}$ with a logit link function $\xi(\cdot)$, where all covariates were generated by the same manner described in Section 4.1, and $\bfgamma=(0.5, 0.5,0.5,0.5,0.5)^T$. In the current situation, $Z_i$ was assumed to be both observable in internal and external studies. With the generated exposure, we further generated potential outcomes ($Y_i(A_i)$)  from two exposure worlds \citep{rosenbaum1983central,robins2000marginal}, i.e., $A=1$ and $A=0$, based on the outcome conditional mean model, i.e., $\xi\{Prob(Y_i=1|\bfX_i, Z_i, A_i=a)\}=\beta_{a0}+\beta_{a1}Z_i+\beta_{a2}X_{1i}+\beta_{a3}X_{2i}+\beta_{a4}X_{3i}$, for $a=0,1$, where $\bfbeta_{a0}=(-0.5, 0.5,-0.5,0.5,-0.5)^T$, and $\bfbeta_{a1}=(0.5, -0.5, 0.5,-0.5,0.5)^T$. The observed outcome was then determined by both potential outcomes and the observed exposure label. Moreover, the true causal model of interest was based on pseudo outcomes and MSM: $\xi\{Prob(Y_i(A_i)=1)\}=\beta_{0}+\beta_{1}A_{i}$, where the true parameter value for the logarithm of causal odds ratio $\beta_{1}$ was calculated by computer simulation using generated potential outcomes under {200000} sample size. 
% \blue{the true parameter value for the logarithm of causal odds ratio $\beta_1$ was calculated by averaging the internal estimates based on $n_1$ over 1000 Monte Carlo runs.}

% The mean function that generates potential $Y_0$ is $\bfX\bfbeta_0$, while the mean function that generates potential $Y_1$ is $\bfX\bfbeta_1$. We let the assignment of treatment follow a binomial distribution with the mean function $\bfX\beta_A$, where the observed outcome equals $Y_1$ if the assignment equals $1$, and vice versa. The $\bfX=(\bfX_1, \bfX_2, \bfX_3)$ is constructed in the same way as in the generalized linear model (Section 4.1) with a potential confounding factor $\bfZ$ that is not observed in the study. The $\bfbeta_1=(-1, 1, -1)^T$, and $\bfbeta_0=(1, -1, 1)^T$, while the assignment parameter is $\bfbeta_A=(1,1,1)^T$. The sample size for internal and external kept the same comparison scheme as in the previous generalized linear model section. 

Moreover, we assumed that the external study considered a conventional logistic regression model, i.e., $\xi\{Prob(Y_i=1|\bfX_i, A_i)\}=\theta_{0}+\theta_1A_{i}+\theta_2Z_i+\theta_{3}X_{1i}+\theta_{4}X_{2i}+\theta_{5}X_{3i}$. The estimated vector of parameter $\hat\bftheta$ and its variance-covariance matrix $\hat\bfV$ were assumed to be available, but not the raw external data. It is worth noting that the external regression model is intrinsically different from the MSM of interest. However, we argue that by integrating information from traditional regression could be still helpful to improve the estimation efficiency in the causal inference model. We evaluated the performance of the new method, denoted by IB\_IPTW, and compared it with the classic MSM estimator based on inverse probability treatment weighting (IPTW) \citep{robins2000marginal} using the PS estimated by logistic regression, in terms of bias, MCSD, and RE, which is the ratio of mean square errors between the IPTW estimator and the proposed IB\_IPTW estimator with data integration. Noted that the GIM-based estimator is not directly applicable in this context.

\begin{table}[]
\caption{Evaluation of the proposed information integration method for estimating the logarithm of causal odds ratio.}\label{causal_tab}
\small
\centering
\renewcommand{\arraystretch}{1.5} 
\begin{tabular}{p{1.2cm}p{1.2cm}p{0.5cm}p{1.2cm}p{1.2cm}p{1.2cm}p{0.5cm}p{1.2cm}p{1.2cm}p{1.2cm}}
\hline
     &       &   & \multicolumn{3}{c}{$n_1=200$}              & \multirow{7}{*}{} & \multicolumn{3}{c}{$n_1=600$}             \\ \cline{1-2} \cline{4-6} \cline{8-10} 
\multicolumn{2}{c}{Ratio of $n_2/n_1$}          &    &  $~~0.75$ & $~~1.5$   & $~~~5$      &              & $~~0.75$ & $~~1.5$   & $~~~5$      \\ \cline{1-2} \cline{4-6} \cline{8-10} 
Bias & IB\_IPTW & & -0.011    & 0.003 & 0.003 &                   & 0.001     & 0.007 & -0.014 \\ %\cline{1-5} \cline{7-9} 
Bias & IPTW    & & -0.012     & -0.005 & 0.003 &                   & 0.001     & -0.002 & 0.001 \\ %\cline{1-5} \cline{7-9} 
MCSD & IB\_IPTW & & 0.341     & 0.304 & 0.271  &                   & 0.183     & 0.169  & 0.143  \\ %\cline{1-5} \cline{7-9} 
MCSD & IPTW    &  & 0.407     & 0.405 & 0.409  &                   & 0.221     & 0.215  & 0.232  \\ %\cline{1-5} \cline{7-9} 
RE   &        &   & 1.423     & 1.774 & 2.282  &                   & 1.452     & 1.602  & 2.611  \\ \hline
\end{tabular}
\end{table}

%\red{add some here, Qu}
%setting: focus on binary outcome case and causal odds ratio parameter (yes, only present results for one parameter, show log odds ratio, not odds ratio \red{good to know, only the odds ratio}). sample size 200 and 600, sample size ratio: 1:0.75, 1:1.5, 1:5. record bias, mcsd, re. In the causal context, the variables used in the $\Psi(\cdot)$ should matter. Let's try two different forms for $\Psi(\cdot)$: 1. include all confounders and A in the $\Psi(\cdot)$ as covariates in the traditional logistic regression; 2. only include A in the $\Psi(\cdot)$ as the covariate in the traditional logistic regression. Also, one for the same x distribution, the other for different x distributions. For causal, you may just use x, no need to consider z in this case.

\textbf{Results}. The results based on $1000$ Monte Carlo runs are summarized in Table \ref{causal_tab}, in different settings of internal sample size $n_1=200,~600$ and sample size ratio $n_2/n_1=0.75,~1.5,~5$. We observed that both IB\_IPTW and IPTW estimators had smaller and closer to zero bias as sample size increased. Compared to the IPTW estimator, the IB\_IPTW estimator showed smaller MCSD and larger than one RE. These results provide valuable numerical evidence supporting our statement that integrating information from the traditional regression model based on the external data could be still helpful to improve the estimation efficiency in the casual model based on the internal data. 

\section{Discussion}
\label{sec:5}
We have provided a selective review of recent and advanced information integration methods via empirical likelihood. Moreover, we provided a new and possibly promising direction to integrate information from a broad context. Compared to existing methods, this new method is computationally more convenient, numerically more stable, and able to integrate summary information from a model that is very different from the main model used for internal analysis. In addition to the simple setting described in Section \ref{Method Framework}, the new framework can be extended to handle more complex settings by adapting techniques described in Section \ref{Various extensions}: for example, if one has observed the issue of heterogeneous covariate distributions between internal and external data, we may consider adopting the technique of semiparametric density ratio model \citep{sheng2022synthesizing, cheng2023semiparametric,huang2023simultaneous} described in (\ref{density ratio model}) into the proposed constraint (\ref{constrain with variance}); if one has concern about heterogeneous conditional outcome distributions, we may impose a bias term assisted by the technique of penalty, similar to the formula in  described in (\ref{bias term}), to alleviate potential bias introduced to the internal estimation; when we have summary information from multiple external data, we may change the joint log-likelihood function in (\ref{Vtheta}) to $\sum_{i=1}^{n_1} log(p_i) - \sum_{k=2}^Kn_k(\hat\bftheta_k-{\bftheta})^T({\hat\bfV}_k)^{-1} (\hat\bftheta_k-{\bftheta})/2$, with $K\geq 2$. Extensive studies are needed to evaluate their validity and utility.

Furthermore, the new framework holds potential for application in various other statistical contexts, including survival analysis, longitudinal data analysis, and analysis of heterogeneous treatment effects, among others, all of which warrant thorough investigation. In terms of applications, the new method has the potential to be applied across diverse disciplines, including the integration of information from multi-center clinical trials, electronic health records from multiple hospitals, and cohort studies from different consortia. In summary, in the era of big data, the authors believe that the framework of information integration and the new idea proposed in this article are poised to play pivotal roles.

% \begin{figure}[b]
% \sidecaption
% % Use the relevant command for your figure-insertion program
% % to insert the figure file.
% % For example, with the graphicx style use
% \includegraphics[scale=.65]{figure}
% %
% % If no graphics program available, insert a blank space i.e. use
% %\picplace{5cm}{2cm} % Give the correct figure height and width in cm
% %
% \caption{If the width of the figure is less than 7.8 cm use the \texttt{sidecapion} command to flush the caption on the left side of the page. If the figure is positioned at the top of the page, align the sidecaption with the top of the figure -- to achieve this you simply need to use the optional argument \texttt{[t]} with the \texttt{sidecaption} command}
% \label{fig:1}       % Give a unique label
% \end{figure}

%
% \begin{acknowledgement}
% If you want to include acknowledgments of assistance and the like at the end of an individual chapter please use the \verb|acknowledgement| environment -- it will automatically render Springer's preferred layout.
% \end{acknowledgement}
%
% \section*{Appendix}
% \addcontentsline{toc}{section}{Appendix}
% %
% %
% When placed at the end of a chapter or contribution (as opposed to at the end of the book), the numbering of tables, figures, and equations in the appendix section continues on from that in the main text. Hence please \textit{do not} use the \verb|appendix| command when writing an appendix at the end of your chapter or contribution. If there is only one the appendix is designated ``Appendix'', or ``Appendix 1'', or ``Appendix 2'', etc. if there is more than one.

% \begin{equation}
% a \times b = c
% \end{equation}

\bibliographystyle{agsm}
\bibliography{Bibliography-MM-MC}

\newpage

\end{document}